\documentclass[%
preprint,
 amsmath,amssymb,
 aps,
]{revtex4-1}
\usepackage{setspace}

\usepackage{graphicx}% Include figure files
\usepackage{dcolumn}% Align table columns on decimal point
\usepackage{bm}% bold math
\usepackage{amsmath}
\newcommand\numberthis{\addtocounter{equation}{1}\tag{\theequation}}
\usepackage{float}
\usepackage{epsfig}

\newcommand{\bra}[1]{\ensuremath{\left\langle#1\right|}}
\newcommand{\ket}[1]{\ensuremath{\left|#1\right\rangle}}

\begin{document}

%\preprint{APS/123-QED}

\title{Weak Classicality in Coupled $N$-Level Quantum Systems}% Force line breaks with \\

\author{Radha Pyari Sandhir}%
 \email{radha.pyari@gmail.com}
\affiliation{%
Department of Physics and Computer Science, Dayalbagh Educational Institute, Dayalbagh Rd., Agra, Uttar Pradesh-282005, India\\
}%
\author{ V. Ravishankar}%
 \email{vravi@physics.iitd.ac.in}
\affiliation{%
 Department of Physics, Indian Institute of Technology Delhi, New Delhi-110016, India\\
}%


\begin{abstract}

\singlespacing
The notion of `weak classical limit' for coupled $N-$level quantum systems as  $N \rightarrow \infty$ is introduced  to understand the precise sense in which one attains classicality.  There exist proofs that a  system becomes classical at large $N$ \cite{Yaffe82,Ghir86}. On the other hand, it is known that non-locality and entanglement, the two hallmarks of non-classicality, thrive even as $N \rightarrow \infty$. We reconcile these results in this paper by showing that so called classicality is not so much  an inherent property of the system, as it is a consequence  of limited experimental resources.  Our focus is largely on non-locality, for which we study the Bell-CHSH and CGLMP inequalities for $N$-level systems.
\begin{description}
\item[PACS numbers]
03.67.-a, 03.65.Ud, 03.65.Ta
\end{description}
\end{abstract}

\pacs{Valid PACS appear here}% PACS, the Physics and Astronomy
\maketitle

%\tableofcontents

\section{Introduction}\label{sect:intro}

\singlespacing
%Secure key generation is an abiding interest in the field of cryptography due to the unbreakability of the one-time pad technique. Quantum information theory has found a place in the security enhancement of key generation techniques due to the quantum properties such as superposition and entanglement. 
The behaviour of quantum systems in the so called large $N$ limit, where $N$ denotes  the dimension of the Hilbert space for some degree of freedom,   holds an abiding interest. This is in view of  the proofs  that a system becomes classical \cite{Yaffe82,Ghir86} as $N \rightarrow \infty$. In the context of spin, e.g., these proofs  exploit an essential property of spin coherent states.   Though spin coherent states are over complete at any finite value of $N$, they  become orthonormal as $N\rightarrow \infty$, making them isomorphic to the classical phase space. This delicate limit has received renewed interest  in studies on non-locality and entanglement in higher dimensional bipartite systems \cite{Merm80,Peres92,Arde91,Arde92,Kasz00}. Contrary to earlier expectations \cite{Merm80}, it was realised  that non-locality and entanglement  continue to thrive even as $N \rightarrow \infty$ \cite{Peres92,Arde91,Arde92,Kasz00}. These results also hold for the more recently proposed inequalities for testing non-locality in higher dimensional systems \cite{Kasz00,Coll02,FuLi04}.    A reconciliation between the theorem mentioned above and these results is in order,  if only for the sake of avoiding inconsistent interpretations. We show that this can be accomplished  through the notion of weak classicality, for which we consider a coupled system of two $N$-level systems.  Thereby this paper also serves a pedagogic purpose.
 
 The paper is organised as follows. As the simplest nontrivial subgroup of $SU(N)$, $SU(2)$ embodies the idea of moments of observables in the $N$-level system. This is because the generators of $SU(N)$ can be constructed from the generators of $SU(2)$ and their higher order tensor moments. We thus demonstrate that observables that yield experimental signatures for both non-locality and entanglement necessarily involve contributions from the highest order tensor moments (Section~\ref{sect:weak}). In this context, an $N \times N$ system is equivalent to a spin $s \times s$ system, and we take $N= 2s+1$.  To substantiate the qualitative observations, we perform quantitative analyses involving (i) polynomial observables in spin operators (Section~\ref{sect:spin_obs}) , and (ii)  projective measurements, which are easily related to actual experimental set-ups (Section~\ref{sect:projops}).  We  proceed to show that the so-called classical limit emerges when  there are  limited experimental resources,  causing an inability to measure  high order moments, and that it is {\it not} an inherent property of the system itself. 
  In view of the well known results obtained in \cite{Brau92,Pope92}, we study the cases of entanglement and non-locality separately.

 \section{Weak classical limit: algebraic analysis} 
 \label{sect:weak}
 \subsection{Non-locality and the weak classical limit}
 A physical system is said to be governed by classical laws if it can be modelled by a theory of local hidden variables \cite{BellInequalities,CHSHOrig}. A violation of the Bell-CHSH inequality is a sufficient, but not necessary signal of the non-locality of a state (see \cite{Brun14} and references cited therein).
 
The family of states that violate the Bell-CHSH inequality includes pure  as well as mixed states, with $N=2$ being the only exception. In fact, pseudo-Bell states, belonging to  all $2 \times 2$ subspaces of $N \times N$ systems,  violate the Bell-CHSH maximally, even though their relative entanglement with respect to the fully entangled state can be negligible. Furthermore, incoherent sums of the pseudo-Bell states belonging to orthogonal subspaces also exhibit maximal violation, even if they are highly mixed \cite{Brau92,Pope92}. 

Let the basis states in the respective Hilbert spaces of the two subsystems be labelled by  $m_i,~\mu_j = -s, -(s-1), \cdots, s-1, s$, in the conventional spin basis. The first example that is of  interest to us is the pseudo two qubit  subspace  spanned  by $ \{\ket{m_i}, \ket{m_j}; \ket{\mu_k}, \ket{\mu_l}\}$ for some fixed values of the subscripts. The subspace supports  states that violate  Bell-CHSH inequality maximally, which are further related to each other by local transformations in the subspace. The observables that reveal  non-locality are simply the Pauli operators in the subspaces.

Simple though this theoretical construction is, isolation   of states in two-dimensional subspaces  is nontrivial experimentally, requiring effective hamiltonians that operate almost entirely in the subspace.  Generically, spin observables span the full space, and measuring mean values  and  their first few moments is  easier, and more feasible.  To see this, consider a Pauli operator, say
\begin{equation}
A = \vert m_i\rangle\langle m_j\vert + \vert m_j\rangle\langle m_i\vert
\end{equation}

defined for the first subsystem. We wish to determine the overlap, $Tr[AT^k_q(\vec{S})]$, of $A$ with the irreducible  tensor operators $T^k_q(\vec{S})$, defined by
 \begin{equation}
 T^k_q(\vec{S}) = C_k (\vec{S}\cdot \vec{\nabla})^k r^k Y_{k,q}(\theta,~\phi) 
 \end{equation}
 where  $\vec{S}$ is the spin operator, and Y$_{k,q}(\theta,~\phi)$, are the spherical harmonics.  The constant $C_k$ is free, and is usually  fixed by setting the  matrix elements to be 
 \begin{equation}
 \langle m_2 \vert T^k_q(\vec{S}) \vert  m_1\rangle = C(sks; m_1qm_2)\sqrt{2k+1}
 \end{equation}
 in terms of the Clebsch-Gordan coefficients. The tensor of rank $k$ yields moments of order $k$, of the spin operators, with $0 \le k \le 2s$. By Wigner-Eckart theorem, the overlap of the Pauli operator $A$ with $T_{q}^k$ thrives even as $k\rightarrow 2s$ so long as $q= \pm \vert m_i - m_j\vert$, and therefore measurements of lower order tensors would not yield the correct expectation value of $A$, rendering the verification of the Bell inequality suspect.  

This is essentially the crux of the argument. Determination of correlations involving direct products of such Pauli operators entails measurements of correlations involving tensor correlations of {\it all} orders,
which would involve varied experimental techniques.
 Thus, if one has the wherewithal to measure moments only  upto a rank $k << 2s$,  it would well nigh be impossible to decide if a state with a spin $s$ is non-local.

 \subsection{Entanglement  and the weak classical limit}\label{sect:entanglement}
 Entanglement and Bell-CHSH non-locality are relative monotones only for pure two-qubit states. We now argue, qualitatively, how weak classicality works with higher dimensional entanglement. For that, it is sufficient to consider a pure state for an $N \times N$ system.
 
 The condition for pure state entanglement is that either of the reduced states have nonzero entropy. In particular,   the reduced states of a fully entangled state  are completely mixed. A verification of this entails complete tomography, which, again,  necessarily involves ascertaining that all the moments, including the highest order $k=N-1$, vanish \cite{Niel11}.  

On the other hand, if the state is separable, the state of the subsystem is pure. A verification of purity again requires the measurement of $t^{2s}_q$, that is, the tensor components.  A spin $s$ state has the resolution
\begin{equation}
\rho=\frac{1}{N}\Big\{1+ \sum_{k=1}^{2s} \sum_{q=-k}^{k} (-1)^q t_{-q}^{k} T_{q}^{k}(\vec{S})\Big\},
\end{equation}
and  it is known  that a pure state is determined, except for residual discrete ambiguities, by the tensor components $t^{2s}_q$ \cite{Ravi86,Ravi87}.

In short, even with ideal measurements, quantumness may not be revealed if the experimental resources are limited in the sense described above. This conclusion complements other results on classical limits that invoke decoherence \cite{Zure91,Brune96,Zure03} or coarse grained measurements \cite{Kofl07}, and of the experimental resolution of adjacent states in a Stern-Gerlach set-up \cite{Peres92}. In all these cases, the classical limit is not seen as an inherent property of the system, but as arising from extraneous exigencies, either because of environment, or precision, or lack of resources. Any of them may mask the quantumness of the state. 

We now buttress  the qualitative arguments by two quantitative analyses that focus on non-locality. We show that, generally, violations of the Bell-CHSH inequality improves, as  the observables have terms that are of increasingly higher orders in spin operators. Almost invariably, these observables have supports in  proper  subspaces of the $N$-dimensional Hilbert space. As mentioned, the Bell-CHSH inequality is violated maximally when the support is two-dimensional.
Keeping this in mind, we consider, 1) observables that are polynomial in $SU(2)$ generators and 2) observables based on projective measurements.   Before we delve into either of the two analyses, we address  the experimental setting.

 \section{Quantitative analysis: the setting}
For our purposes, it is sufficient to consider the completely entangled state (invariant under global $SU(N)$),
 \begin{equation}
 |0_s> =\frac{1}{\sqrt{2s+1}}\sum_{m=-s}^{s}(-1)^{m+s}|m,-m>.
 \end{equation}  
 To check the robustness of the results,  we  employ the CGLMP formulation \cite{Coll02},  as well as the Bell-CHSH functions arising from a large family of correlations. Note that CGLMP  yields non-locality criterion that is  dimension dependent. We do not discuss violations coming from proper subspaces since they have essentially been considered in the previous section.
 
 Consider, first, the Bell-CHSH function
 \begin{equation}
\mathcal{B} \equiv |\mathcal{C}(a, b) - \mathcal{C}(a, b')|+|\mathcal{C}(a',b)+\mathcal{C}(a',b')|,
\label{eq: bellinequality}
\end{equation}
which signals non-locality whenever  $ 2 < {\cal B} \le 2\sqrt{2} $. Furthermore, we employ the standard detector configuration geometry defined by
\begin{equation}
\theta_{ab}=\theta_{a'b}=\theta_{a'b'}=\frac{\theta_{ab'}}{3},
\label{eq:angles}
\end{equation}
The precise meaning of $\theta$ will become clear from the examples discussed below.
This prescription is not {\it ad hoc} for two reasons. An extensive study, partly analytical and partly numerical has shown this geometry to be optimal \cite{Mend05}. Our own computation confirms that maximal violations occur at this geometry for higher spins as well. The parameter space is thus rendered one dimensional.

\section{Observables that are polynomial in $SU(2)$ generators}\label{sect:spin_obs}

Our quantitative studies begin with correlations arising from polynomial observables in spin operators of equal degree. We make use of the notation $\mathcal{C}^s_k$, where $k$ is the order of the correlation and $s$ is the spin of the state.

\subsection{Linear correlations}\label{sect:vector}

Recall that the state is a spin singlet state, invariant under global $SU(N)$ transformations. Consider
\begin{equation}
\mathcal{C}_{1}^s(\hat{a}, \hat{b}) =  \langle 0_s\vert (\frac{\vec{S}}{s}\cdot \hat{a}) (\frac{\vec{S}}{s}\cdot \hat{b}) \vert 0_s \rangle
\label{eq:dipole}
\end{equation}
in terms of the normalised spin operators $\frac{\vec{S}}{s}$, which are the standard angular momentum operators,  respecting the unit norm bound, $\vert Tr\{\rho^s \frac{\vec{S}}{s} \}\vert \le 1$.
Let $\cos \theta_{ab} \equiv \hat{a}\cdot \hat{b}$. Exploiting the isotropy of the state,  we find
\begin{equation}
\mathcal{C}^s_1(\hat{a}, \hat{b}) = -\frac{s+1}{3s} \cos \theta_{ab} \equiv -\frac{s+1}{3s}(\hat{a}\cdot \hat{b}).
\label{eq:linear_corr}
\end{equation}
where $\mathcal{C}_1^{\frac{1}{2}}=-(\hat{a}\cdot \hat{b})$, is the well known spin-spin correlation for qubits in the singlet state.  Note that the spin dependent scaling factor, which is also inherited by the corresponding Bell-CHSH function ${\cal B}^s_1$,  decreases rapidly with $s$, achieving an asymptotic value  of $\frac{1}{3}$. It falls by a factor 2/3 for $s=1$. Since the maximum violation for two qubit case is $2\sqrt{2}$, it follows that the Bell-CHSH inequality is respected by the correlation for spins $s \ge 1$. One is therefore obliged to look at  correlations that involve higher orders in spin.

\subsection{Biquadratic correlations}\label{sect:generic}
Consider the generic quadratic observable
\begin{equation}
O_A  = C_2 \Bigg(\frac{\vec{S}\cdot \hat{a}}{s}\Bigg)^2 + C_1 \Bigg(\frac{\vec{S}\cdot \hat{a}}{s}\Bigg) + C_0
\label{eq:quadg}
\end{equation}
with the proviso that its expectation values be bounded by unit norm. We fix  the optimal values of the constants $C_i$  through a numerical search. As a check, we find that the numerical results agree with the analytic results, which are easily available for $s=\frac{1}{2},~\frac{3}{2}$. 

The unit norm bound on the coefficients $C_i$ in Eq. \ref{eq:quadg} simply translates to the  set of $2s+1$ constraints (in each $m$ value):
\begin{equation}
-1\leq \frac{C_2 m^2}{s^2}+\frac{C_1 m}{s}+C_0\leq 1 
\label{eq:constraints_quad}
\end{equation}
 in the three parameters. 
 
 The  data points for the coefficients are numerically generated by selecting the bounds [-5,5] for each coefficient here and everywhere. These bounds were selected on the basis of observation, as being ample enough to address the question under study. Each of the bounds  yields a two-dimensional plane in the $N-$dimensional space, and the intersections of the planes gives us the vertices of a polyhedron within and on which the coefficients $C_i$ are constrained to lie. The additional requirement that the observables admit their maximum value restricts our search to identifying one of the vertices. We present the results for various spins in the following subsections:
 
\subsubsection{$s=1$}
We find that the maximum violation occurs when $C_2=2;~C_1=0;~C_0=-1$. The numerical search confirms that the maximum violation occurs in the planar geometry, with ${\cal B}_{max}=2.55\pm\epsilon$, matching the result found in \cite{Peres92}.
 The maximum violation, pegged at the value 2.55,  falls short of the maximum allowed value, $2\sqrt{2}$, which is realized for a two qubit system -- notwithstanding the fact that we are dealing with a completely entangled pure state.  This result is not surprising since fully entangled states with $N$ odd fail to violate Bell inequality maximally \cite{Sandh17}.
   
\subsubsection{$s \ge \frac{3}{2}$}
 Consider $s=\frac{3}{2}$ first.  The optimal  values of the coefficients are found to be   $C_2=1, C_1=0,C_0=-1.25$, yielding the observable
\begin{equation}
O_A=(\vec{S}\cdot \hat{a})^2-\frac{5}{4}
\label{eq:gen_quad_op_s1p5}
\end{equation}
which can be recast into the elegant form    
\begin{equation}
O_A (\hat{a}) = \Pi_{3/2}(\hat{a})  + \Pi_{-3/2}(\hat{a}) - \Pi_{1/2}(\hat{a})- \Pi_{1/2}(\hat{a}) 
\label{eq:gen_quad_s1p5_op}
\end{equation}
in terms of the projection operators $\Pi_m \equiv \vert m \rangle \langle m \vert$ along the quantization axis $\hat{a}$,
with a similar expression for $O_B$. Note that expectation values of $O_{A,B}$ span the full range 
$[-1,+1]$. The correlation has the compact form:
\begin{equation}
\mathcal{C}_2^{3/2} = P_2(\cos\theta)
\end{equation}
which has its support in $[-\frac{1}{2}, 1]$, larger than the one obtained for $s=1$. The maximum value of the Bell-CHSH function  is 2.62, which is larger than the violation for $s=1$. Incidentally, 
note that the coincidence count rates to be measured are clear from the very expression for  the observables given in Eq. \ref{eq:gen_quad_s1p5_op}. The increase in the violation from $s=1$ to $s=\frac{3}{2}$ is anomalous. In fact, we find that the biquadratic correlation fails to violate Bell-CHSH inequality everywhere in the parameter space if $s \ge 2$.

It is getting clear from the studies so far that violation of the Bell-CHSH inequality requires correlations of observables that are of maximal order in the spin variables, $k \simeq 2s$. To further substantiate this finding,   and also to draw more quantitative conclusions for comparison with similar findings \cite{Merm80,Peres92,Gis92,Wodk95,Bana98}, we consider two more cases -- generic  observables of degree three, and a specific observable of degree four.  This necessitates a discussion of $s=2$ states as well. We address the cubic case first, and conduct a global search in the parameter space.

\subsection{Bicubic correlations }
The generic form of the observable, say, for the subsystem $A$, is given by
\begin{align}
O_A=C_3\Bigg(\frac{\vec{S}\cdot \hat{a}}{s}\Bigg)^3+C_2\Bigg(\frac{\vec{S}\cdot \hat{a}}{s}\Bigg)^2+C_1\Bigg(\frac{\vec{S}\cdot \hat{a}}{s}\Bigg)+C_0.
\end{align}
The usual requirement that $O_A$ be bounded by unit norm leads to the set of  $2s+1$ constraints
\begin{equation}
 -1 \le \frac{C_3 m^3}{s^3}+\frac{C_2 m^2}{s^2}+\frac{C_1 m}{s}+C_0 \le1~ \forall m \in [-s,s].
\label{eq:constraints_cubic}
 \end{equation}
As before, the coefficient data was generated by selecting the bounds [-5,5]. The search in the parameter space yields a maxima in the violation of the Bell-CHSH inequality in two orthogonal subspaces,  of even and odd parity in spin observables. The even parity case was discussed in the previous section.  We address the complementary space. 
\subsubsection{$s=\frac{3}{2}$}
A numerical search yields the maximum violation for the correlation when 
$C_3=\frac{4}{3},C_1=-\frac{7}{3},C_2=C_0=0$, corresponding to the observable

\begin{equation}
O_A=\frac{4}{3}(\vec{S}\cdot \hat{a})^3-\frac{7}{3}(\vec{S}\cdot \hat{a})
\label{eq:numericalop_1p5}
\end{equation}

For this optimal observable, the maximum violation is pegged at $\mathcal{B}^{\frac{3}{2}}_3=2.47$.

\subsubsection{$s \ge 2$}
We conclude  the discussion on cubic correlations with a brief discussion of higher spins. The state with $s=2$ shows a mild violation, with a maximum value  of the Bell-CHSH function given by $\mathcal{B}^{2}_3=2.03$ for the configuration $C_3=0.5,C_1=-\frac{3}{2},C_2=C_0=0$.  The Bell-CHSH inequality is respected for the cubic correlations involving all spins $s\geq\frac{5}{2}$.

\subsection{Biquartic correlations for $s =2$}\label{sect:4pole}
In this last of the illustrations, we  consider a specific quartic observable for $s=2$, which is chosen to be
\begin{equation}
O_A=2(\vec{S}\cdot \hat{a})^4-1.
\end{equation}
This clearly spans the full range $[-1,+1]$. This observable is nontrivial for $s \ge 2$, and we restrict ourselves to spin 2 here. 
A straightforward evaluation yields the correlator to be  
\begin{equation}
\mathcal{C}^2_4=-\frac{7}{10}+\frac{45}{32}\sin^2\theta\cos^2\theta+\frac{39}{640}\sin^4\theta+\frac{51}{32}\cos^4\theta
\label{eq:quartic_corr}
\end{equation}

The maximum value attained by the Bell-CHSH function is  $\mathcal{B}^2_{4\max}=2.371$. This may be combined with the violation seen for the cubic correlator, as the spin observables act in complementary spaces.

\subsection{Summary of results}\label{sect:4pole}

The examples considered above suggest strongly that 
\begin{enumerate}
\item  Non-locality in $N$-level systems will be seen if the observables employed are of degree $N -1$,  with a small leeway for lower ordered observables close to $N$. 
\item  The magnitude of violation does not seem to have significant diminution with increasing $N$. 
\item This does not, however, shed light on the complete extent of violation: spaces of complimentary orders hold violations as well, and so, a complete family of observables appears to be needed to be taken into account. 
\item Finally, in spite of optimization, and in spite of employing the most non-local state, the violation fails to reach the maximum allowed value, $2\sqrt{2}$, for polynomials in spin observables.
\end{enumerate}

The results are collated  and displayed in table~\ref{tab:summary}.

\begin{table}[H]
\centering
\caption{Summary of Results}
\label{tab:summary}% Give a unique label
\begin{tabular}{|c|c|c|c|} 
 \hline
 \textbf{$s$} & \textbf{$\mathcal{O}(O_{A,B})$}  & \textbf{$O_A$}  & \textbf{$\mathcal{B}_\text{max}$} \\
 $\frac{1}{2}$  & 1 & $(\vec{\sigma}\cdot \hat{a})$ & $2\sqrt{2}$ \\
 1 & 2 & $2(\vec{S}\cdot \hat{a})^2-1$ & 2.55  \\
 $\frac{3}{2}$ & 2 & $(\vec{S}\cdot \hat{a})^2-\frac{5}{4}$ & 2.62 \\
  $\frac{3}{2}$ & 3 & $\frac{4}{3}(\vec{S}\cdot\hat{a})^3-\frac{7}{3}(\vec{S}\cdot\hat{a})$ & 2.47 \\
 2 & 3 & $\frac{1}{2}(\vec{S}\cdot \hat{a})^3-\frac{3}{2}(\vec{S}\cdot \hat{a})$ & 2.03 \\
 2 & 4 & $2(\vec{S}\cdot \hat{a})^4-1$ & 2.37 \\
 \hline
\end{tabular}
\begin{center}{\emph{Maximal violation ($\mathcal{B}_\text{max}\pm\epsilon$ ) of the Bell-CHSH inequality obtained for the spin-$s$ singlet state with polynomial observables $O$ in spin operators of order $\mathcal{O}$.}}  \end{center}
% Or use
%\vspace*{5cm}  % with the correct table height
\end{table}

\section{Observables based on projective measurements} \label{sect:projops}
This section extends the results of the previous section to the larger class of observables that are  based on projective measurements, a form in which any generic observable can be constructed. Observables that are constructed by projective measurements translate operationally to the lab as Stern-Gerlach-like measurements, which will be useful for future experimental work. Additionally, these observables are of the same form employed by Peres \cite{Peres92}, whose observables are a subset of those considered here.

\begin{align*}
O(\hat{q})=\sum (-1)^{f_m} |m><m|  \leftrightarrow 
\mathbb{P} = \sum_{m=-s}^{m=+s}f_m 2^{m-s}; \numberthis
\label{eq:obs}
\end{align*}
where $f_m =0,1$.  Here $\hat{q}$ refers to the choice of the quantization axis and $f_m$  takes values $\{0,1\}$. The sequence $\{f_m\}$ -  henceforth called the `parity bit array' -  characterizes the observable. In turn,  each parity bit array admits a unique integer representation
$\mathbb{P}$,  which we may dub  as the  `parity bit integer'. We employ this notation throughout. In short,  we have shortlisted a set of $2^N$ observables for each subsystem.  Note that the result of a measurement of  $O(\hat{q})$ can be easily expressed in terms  of appropriate count rates for various quantum numbers $m$. Further note that each observable partitions the Hilbert space to a direct sum of two orthogonal subspaces, ${\cal H_N} = {\cal H}_k \oplus {\cal H}_{N-k}; ~k= 0,\cdots \frac{[N]}{2}$.

\subsection{The multilinears}
Since the subsystems are treated on an equal footing in the state, we choose  identical observables  for $A$ and $B$, both labelled by the same $\mathbb{P}$, which also characterises the correlator $O_AO_B$. Thus we are equipped with  a set of $2^N$ multilinears of which the one  proposed by Peres \cite{Peres92} emerges as a special  example.

To identify the parameter space for the Bell-CHSH function in Eq.~\ref{eq: bellinequality},  we note that though the isotropic state is invariant under global $SU(N)$, for simplicity, we restrict our observables and multilinears to those that pertain to the smaller group of transformations,  {\it viz} $SU(2)$. This makes the visualization of multilinears simpler,  and  experimental verifications easier.   

With this restriction, the two observables differ only  in their quantization axes, which we denote by $\hat{a},\hat{b}$ respectively. We denote the corresponding spin bases by $\{\vert m\rangle\}$ and $\{\vert n\rangle\}$. The two bases are related via the $N$-dimensional irreducible representation (irrep) of $SU(2)$. Explicitly,
\begin{equation}
\vert n\rangle =  \sum_m D^s_{m,n}(\phi, \theta, \psi) \vert m \rangle
\end{equation}
where $D^s_{m,n}$ are the Wigner matrices. The correlation function, given by 
\begin{equation}
C(\hat{a}, \hat{b})= \langle \Psi \vert O_A(\hat{a})O_B(\hat{b}) \vert \Psi \rangle, 
\label{corr}
\end{equation}
 is  invariant under rotations, and depends only on $\cos\theta \equiv \hat{a}\cdot \hat{b}$. It is found to be
\begin{equation}
C(\hat{a}, \hat{b}) = \frac{1}{2s+1} \sum_{n,m}(-1)^{f_n+f_m}|d^{s}_{-m,n}(\theta)|^2
\label{eq:correxp}
\end{equation}
where the $d$ matrix represents a rotation about the $y$ axis by an angle $\theta$.

\subsection{Computational complexity}
We have at hand a set of $2^N$ multilinears, of which the one corresponding to $f_m=1$ for all $m$ is trivial. Furthermore, an overall parity operation on any observable leaves the Bell-CHSH function intact, leading to $2^{N-1} -1$ independent multilinears. It is possible that two physically distinct multilinears $\mathbb{P}_i,~\mathbb{P}_j$  may yield the same expectation value $C(\hat{a},\hat{b})$.  Before we identify distinct correlation functions, we quickly examine the computational issues involved in evaluating the elements of the irreps of $SU(2)$, the Wigner matrices.
\subsection*{The structure of the Wigner-$D$ matrix}
 The  elements of the $D$ matrix, occurring in  Eq.~\ref{eq:correxp},  do admit an analytic expression for any spin, but their  form becomes increasingly complicated to evaluate with increasing dimensions. Probing non-locality  analytically also becomes increasingly frustrating, barring some exceptional cases \cite{Peres92}. We employ quasi numerical techniques to evaluate the multilinear as well as the Bell-CHSH function.

\begin{table}[H]
\centering
\caption{Number of Distinct Multilinears for Spin-$s$}
\label{tab:unique_corr}% Give a unique label
\begin{tabular}{|c|c|c|c|c|} 
 \hline
$s$ & $D(\mathcal{H})$ &$N_D$ & $\mathcal{V}^s_\mathcal{C}$ & $\mathcal{N}^s_\mathcal{C}$ \\
\hline
1 & 9 & 2 & 3 & 2 \\ 
3/2 &16 & 4 & 7 & 5 \\
 2 & 25 &6 & 15 & 9 \\ 
 $5/2$ & 36 & 9 & 31 & 19 \\
 3 & 49 & 12 & 63 & 35 \\
  $7/2$ & 64&16 & 127 & 71 \\
 4 & 81&20 & 255 & 135 \\
 $9/2$ & 100&25 & 511 & 271 \\
 5 & 121& 30 & 1023 & 527 \\
 $11/2$ &144& 36 & 2047& 1055 \\
  6 & 169& 42 & 4095 & 2079 \\
   $13/2$ &196& 49 & 8191 & 4159 \\
7& 225 &56 & 16383 & 8255 \\
  \hline
\end{tabular}
\begin{center}{\emph{$s$ is the spin, $D(\mathcal{H})$ is the dimension of the Hilbert space, $N_D$ is number of unique $d$-matrix elements, $\mathcal{V}^s_\mathcal{C}$ is  number of distinct parity bit integers, and $\mathcal{N}_\mathcal{C}$ is number of distinct multilinears.}}  \end{center}
% Or use
%\vspace*{5cm}  % with the correct table height
\end{table}

Being irreps of $SU(2)$, the Wigner matrices respect the following symmetries:  
\begin{equation}
|d_{m_1m_2}|^2=|d_{-m_1-m_2}|^2=|d_{m_2m_1}|^2=|d_{-m_2-m_1}|^2
\label{eq:symmetry}
\end{equation}
which essentially obviates the need to evaluate 75\% of the matrix elements in $d^s_{m,n}$. The number of independent matrix elements is given by $N_D=s(s+1)$ for integer $s$ and $N_D=(s+\frac{1}{2})^2$ for half odd integer $s$.  
Also, the sum of the diagonal elements is common to all the multilinears. This mitigates the computational difficulty marginally,  and one  is still left with the task of executing a sum   over terms of $O(N!)$ for each matrix element. 

The symmetries also allow different multilinears to yield the same expectation value in Eq. \ref{eq:correxp}. 
Table~\ref{tab:unique_corr} illustrates the points made, where we display the details  upto $s=7$ ($N= 15$). Of particular interest are the last two columns which enumerate the total number of multilinears $\mathcal{V}^s_\mathcal{C}$,  and the number of distinct correlation functions,  $\mathcal{N}^s_\mathcal{C}$. 
  One may surmise that  $\mathcal{V}^s_\mathcal{C} \sim \frac{1}{2}\mathcal{N}^s_\mathcal{C}$ which still leaves us with an exponentially large number to reckon with. The results on these distinct multilinears will be presented systematically in the following sections.    Overall, our analysis covers 1397 independent multilinears, of which 732 have distinct functional forms. Each multilinear was found to violate the Bell-CHSH inequality, though not to the extent of the Tsirelson bound, thereby providing a fertile ground for experimentalists to test non-locality.  Thus equipped, we discuss the weak classical limit.
  
The multilinears that show maximum violation are listed in Table \ref{tab:select_bell_results}. 

\begin{table}[H]
\centering
\caption{Maximum Bell-CHSH Violations for Spin $s$}
\label{tab:select_bell_results}% Give a unique label
\begin{tabular}{|c|c|c|c|} 
 \hline
$s$ & $D(\mathcal{H})$ & $\mathbb{P}$ & \textbf{$\mathcal{B}_\text{max}$}   \\
 1  & 9&5 & 2.55     \\
 $\frac{3}{2}$ & 16 & 9 & 2.62 \\
 2 & 25 & 17 &  2.53 \\
 $\frac{5}{2}$ & 36& 54 & 2.56 \\
 3 & 49&77 & 2.51 \\
 4 & 81 & 306 & 2.51 \\
 5 & 121&1212 & 2.51\\
  \hline
\end{tabular}
\begin{center} \emph{$D(\mathcal{H})$ is the dimension of the Hilbert space, $\mathbb{P}$ is the parity bit vector, and $\mathcal{B}_\text{max}$ is the maximum violation across all multilinears for spin $s$. }\end{center}
% Or use
%\vspace*{5cm}  % with the correct table height
\end{table}

 Note that these violations are either greater than or equal to the corresponding violations in Table~\ref{tab:summary}.  Therefore, observables based on projective measurements appear to be more natural indicators of non-locality than those that are polynomials in $SU(2)$ generators.

\subsection{Overlap with tensor moments}

We now establish the non-vanishing contribution of higher order tensor moments to the observables.
The argument simply mimics the one employed in Section~\ref{sect:weak}). It is also sufficient to consider the family of tensors $T^k_0(\vec{S})$. Consider thus
\begin{eqnarray}
\eta & = &Tr\{O(\hat{q})T^{2s}_{0}(\vec{S})\}  = \sum_m (-1)^{f_m}\langle s,m \vert T^{2s}_{0}(\vec{S}) \vert s,m\rangle \nonumber \\
& = & \Big\{ \sum_m (-1)^{f_m} C(s ~2s ~s; m~ 0 ~ m) \Big\} \langle s \vert\vert T^{2s} \vert\vert s \rangle,
\end{eqnarray}

We find the maximum overlap of $T_0^k,(k=2s,2s-1,2,1)$ with the family of observables accessible to the spin-$s$ system.  The results are collated in Table~\ref{tab:overlap}.

\begin{table}[H]
\centering
\caption{Overlaps Between Tensor Moments and Projective Observables}
\label{tab:overlap}% Give a unique label
\begin{tabular}{|c| c| c |c |c |c|}
\hline
 s & N & $T^{2s}_0(\vec{S})$ & $T^{2s-1}_0(\vec{S})$ & $T^{2}_0(\vec{S})$ &$T^{1}_0(\vec{S})$ \\ 
 \hline
 2 & 5 & 0.8552 & 0.8485 & 0.9561 & 0.8485 \\
 \hline
 4 & 9 & 0.7521 & 0.7965 & 0.8863 & 0.8606\\
 \hline
 6 & 13 & 0.6908 & 0.7473 &  0.8678 & 0.8634\\
 \hline
\end{tabular}
\begin{center} \emph{Maximum overlap $\eta$ between tensor moment $T^{k}_0(\vec{S})$ and the family of projective observables in Eq.~\ref{eq:obs}  }\end{center}
% Or use
%\vspace*{5cm}  % with the correct table height
\end{table}

It is evident that the degree of overlap of the highest order $2s$ is not only non-vanishing, its contribution is comparable to lower orders.   The contribution only vanishes when $f_m$ takes a constant value; the trivial exception when the projection is the identity operator in ${\cal H}_N$. The matrix elements of these operators are, therefore, strongly dependent on $N$, however large $N$ may be. 

\section{Weak Classicality in the CGLMP prescription}\label{sect:cglmp}

Bell-CHSH is but one way of describing non-locality; there are many other inequivalent prescriptions that have subsequently been proposed. We check the robustness of our conclusions by extending the analysis to the CGLMP prescription \cite{Kasz00,Coll02,FuLi04}. This prescription differs from  Bell-CHSH mainly in dimension dependent inequalities.

\begin{figure}[H]
\centering
\includegraphics[width=0.6\textwidth]{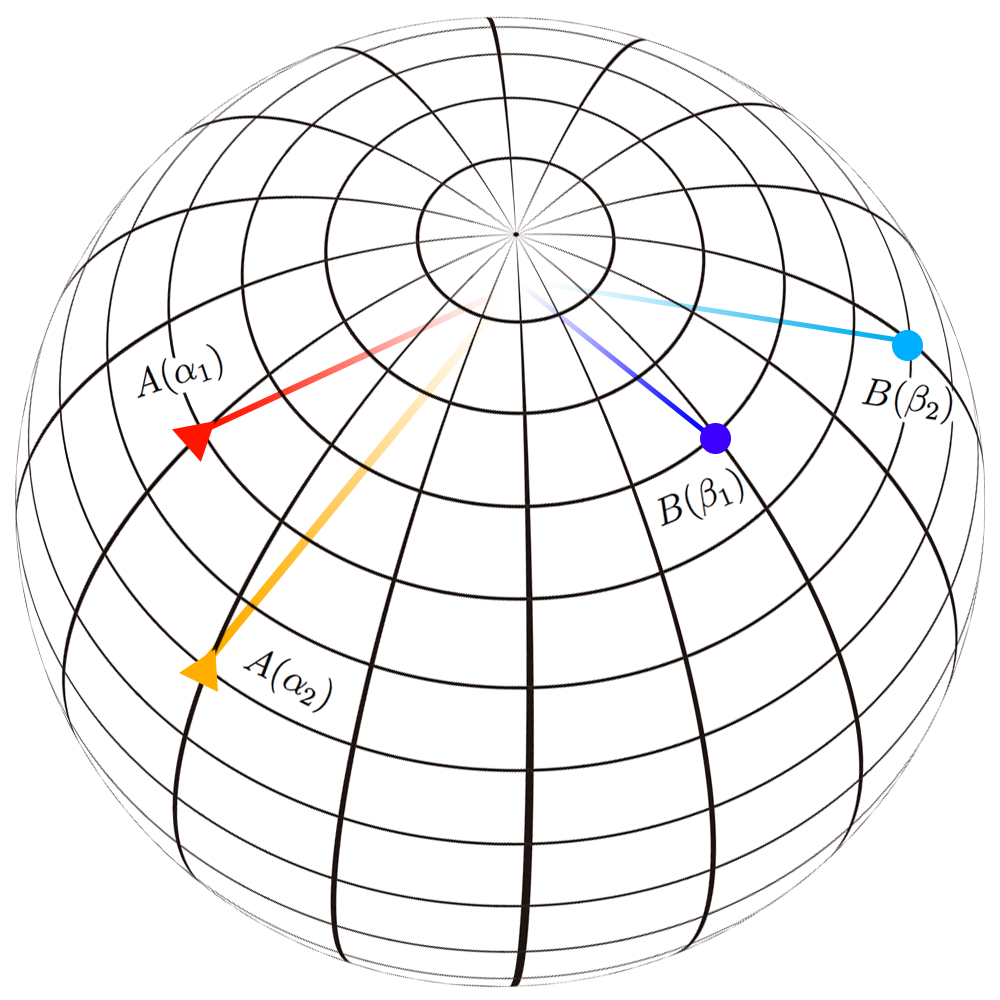}
\caption{Observable schematic for the CGLMP prescription of non-locality.  Variable phases $\alpha_1,\alpha_2, \beta_1, \beta_2$ are fine-tuned to set the pairs of observables in subsystems A and B. They are represented as vectors with intersections on the unit sphere being depicted by triangles and circles respectively.   \emph{(Colour online.)}}
\label{fig:CGLMPoperators}
\end{figure}

The measurement process in this prescription follows the methodology in \cite{Durt01}. As with the Bell-CHSH case, it involves two local observers,  Alice and Bob, each with pairs of dichotomic observables accessible to them. Variable phases $\alpha_i,\beta_i$ are fine tuned to set observables of their choice depending on the measurements they wish to perform (Figure~\ref{fig:CGLMPoperators}). The measurement bases for the observables $A_i$ and $B_i$; $i = 1,2$ are mutually unbiased and of the form:
\begin{eqnarray} \label{eq:basis}
\ket{R}_{A,i} & = & \frac{1}{\sqrt{N}} \sum_{j=0}^{N-1}{exp \Big(i\frac{2 \pi}{N} j (R + \alpha_i) \Big) \ket{j}_A} \nonumber \\
\ket{S}_{B,i} & = & \frac{1}{\sqrt{N}} \sum_{j=0}^{N-1}{exp \Big(i\frac{2 \pi}{N} j (-S + \beta_i) \Big) \ket{j}_B}.
\end{eqnarray}

The observables take the form $A=\sum_R R\ket{R}\bra{R}, B=\sum_S S\ket{S}\bra{S}$, where $R,S$ are respective eigen values. These observables are in general non-zero in their off-diagonals.  Higher order tensor relationships are required to manifest such non-zero off-diagonals, as discussed earlier.  

In the CGLMP prescription, the analog to the Bell-CHSH function given by Eq.~\ref{eq: bellinequality}, for the detection of non-locality is the function:
\begin{align}\label{eq:cglmp_inequality}
&I_N = \sum_{t=0}^{[N/2]-1} \Big(1- \frac{2t}{N-1}\Big) \{  [P(A_1 = B_1 + t) \\  \nonumber
&+ P(B_1 = A_2 +t+1) + P(A_2 = B_2 +t)  \\ \nonumber
& + P(B_2 = A_1 +t)] - [P(A_1 = B_1 -t -1) \\ \nonumber  
&+ P(B_1 = A_2 -t)  + P(A_2 = B_2 -t -1)  \\ \nonumber
&+ P(B_2 = A_1 -t -1) ] \}
\end{align}

where $P(A_i,B_i)$ are joint measurement probabilities for local observables $A_i,B_i$ belonging to the two subsystems. All the observables have distinct integer eigenvalues $0,1, \cdots, N-1$. 

The CGLMP operator is nothing but a linear combination of projectors, with probability $P(A_1 = B_1 + t)$, for instance, determined by projector \\ $\ket{S_1+t}\bra{S_1+t}\otimes \ket{S_1}\bra{S_1}$.

Each term requires a contribution from not only lower ranked, but higher ranked moments as well.  It suffices to show that a single projector of the form $\ket{R}\bra{R}$ has contribution from tensors of all ranks, i.e.,  it suffices to show a non-zero overlap of one projector. We once again consider the overlap between tensor moment $T^k_q$ and the form of the operator.  Since we are dealing with spin moments, the prescription is re-defined in terms of spin: $N=2s+1$, with $j_i \in [-s,s]$.   We look at the projection corresponding to $R$, for an observable defined by phase $\alpha_i$.

\begin{align}
&Tr\bigg[\ket{R}\bra{R} T_q^k\bigg]\\ 
&=\frac{1}{N}\sum_{j_1j_2}\exp{\bigg(\frac{2\pi i}{N}(R+\alpha_i)(j_1-j_2)\bigg)} \bra{j_1}T_q^k\ket{j_2} \numberthis
\label{eq:cglmpoverlap}
\end{align}

where matrix element: \\ $\bra{j_1}T_q^k\ket{j_2} =C(s,k,s;j_2,q,j_1)\bra{s}|T^k_q|\ket{s}$, \\ and $\bra{s}|T^k_q|\ket{s}=\sqrt{2k+1}$.

We consider specific projection operators on 3 and 4-level systems for phases $\alpha=0,\frac{1}{2}$, those used initially in \cite{Coll02}.

\begin{table}[H]
\centering
\caption{Overlaps Between Tensor Moments and a CGLMP Projector}
\label{tab:cglmpoverlap}% Give a unique label
\begin{tabular}{|c| c| c |c |}
\hline
 $s$  & $k=1$ & $k=2$  & $k=3$ \\ 
  \hline
 1 & $\sqrt{2}$ & $0.2\sqrt{15}$ & NA \\
 \hline
 $\frac{3}{2}$& $\approx 1.995$ & $0.4\sqrt{10}$  & $\approx 0.846$  \\
 \hline
\end{tabular}
\begin{center} \emph{Absolute values of the cumulative overlap $\sum_q T^{k}_q$ between tensor moment $T^{k}_q$ and projector, given by Equation~\ref{eq:cglmpoverlap}. Note: overlaps were identical for $\alpha=0,\frac{1}{2}$ and overlaps with all $\ket{R}\bra{R}$ where $R=-s\cdots s$ were identical. }\end{center}
% Or use
%\vspace*{5cm}  % with the correct table height
\end{table}

The overlaps were identical for both phases, indicating that phase does not contribute to the manifestation of these tensor moments.  Moreover, each projector has the same magnitude of contribution for a given tensor operator.  It is clear that tensor moments of order $2s$ contribute to the detection of non-locality. By simply projecting the CGLMP functional operator onto the subspace given by $T^1_q$, one loses information related to tensor moments of higher ranks.   Thus, this dependence is not unique to a particular prescription of non-locality, and true non-locality cannot be revealed without contributions from these higher ordered tensors.

\section{Conclusion}
In conclusion, we have shown, through qualitative and quantitative analyses, that the emergence of classicality in the so called large $N$ limit is valid only in a weak sense, being entirely due to the limited experimental resources available to determine  tensor moments of observables and correlations. The result has been demonstrated for  both entanglement and non-locality, with two different prescriptions for the latter.   In short, the so-called classical limit is not an inherent property of the system.   Consequently, the emergent classicality in higher dimensions is  `weak' in the sense defined and described in this paper.

\section*{Acknowledgements}
We are grateful for our insightful interactions with Konrad Banaszek. RPS thanks the Department of Science and Technology (DST), India, for the Women's Scientist Scheme fellowship [SR/WOS-A/PS-29/2013 (G)].   RPS also thanks the Indian Institute of Technology, Delhi, for providing the resources to pursue this work. 

\bibliographystyle{unsrt}
%\bibliography{Bibliography}

\end{document}